\documentclass[a4paper, 10pt, conference]{ieeeconf}      

\IEEEoverridecommandlockouts                              
\usepackage{graphicx}                                                   
\usepackage{caption}
\usepackage{subcaption}

\usepackage{color, colortbl}
\definecolor{Gray}{gray}{0.9}

\title{\LARGE \bf
Training on the Edge: The why and the how
}

\author{Navjot Kukreja$^{1}$, Alena Shilova$^{2}$, Olivier Beaumont$^{2}$, Jan H\"uckelheim$^{1}$, \\ Nicola Ferrier$^{3}$, Paul Hovland$^{3}$, Gerard Gorman $^{1}$
\thanks{$^{1}$Department of Earth Science and Engineering, Imperial College London
        {\tt\small n.surname@imperial.ac.uk}}%
\thanks{$^{2}$ Inria Bordeaux, France
        {\tt\small name.surname@inria.fr}}%
\thanks{$^{3}$Argonne National Laboratory, IL, USA
        {\tt\small }}%
}

\begin{document}

\maketitle
\thispagestyle{empty}
\pagestyle{empty}

\begin{abstract}

Edge computing is the natural progression from Cloud computing, where, instead of collecting all data and processing it centrally, like in a cloud computing environment, we distribute the computing power and try to do as much processing as possible, close to the source of the data. There are various reasons this model is being adopted quickly, including privacy, and reduced power and bandwidth requirements on the Edge nodes. While it is common to see inference being done on Edge nodes today, it is much less common to do training on the Edge. The reasons for this range from computational limitations, to it not being advantageous in reducing communications between the Edge nodes. In this paper, we explore some scenarios where it is advantageous to do training on the Edge, as well as the use of checkpointing strategies to save memory. 

\end{abstract}

\section{Introduction}
\label{sec:introduction}
Edge computing is a paradigm where computing capability is distributed across a large number of small devices instead of being concentrated in centralized ``cloud'' systems. This places more computing capability closer to either the user, or the source of the data~\cite{garcia2015edge} As it gets cheaper to have additional compute capability in devices like cellphones, cameras and environmental sensors, it becomes viable to do more processing on these devices.   
The ability to do processing at the Edge can be useful for many reasons. By running computations at the Edge, the latency on these computations can be reduced. This can be useful where a quick result is important, e.g. self-driving cars. By decreasing communication, we reduce the bandwidth and power requirements on the edge nodes - and thus the cost of infrastructure setup. The decentralization that comes with having a few hundred devices also increases the reliability of the system since a centralized system is likely to have a single point of failure. Another reason for doing more computation on the Edge is privacy, since in-situ processing avoids sending potentially sensitive information over a network~\cite{paul2014privacy}. For all these reasons, it is becoming commonplace to do inference on machine learning models on the Edge. In Section~\ref{sec:waggle}, we discuss a project that uses this model extensively. 

While it is common to do inference on the Edge today, performing training on the Edge is still not a common paradigm. There are many reasons for this. Firstly, if the information to be learned is relevant to the other Edge nodes, transferring a model update back and forth between the different nodes might introduce excessive communication and increase bandwidth requirements and latency. In this scenario, it might be more efficient to do the training centrally and only transfer the updated model to the Edge nodes.  
Secondly, machine learning models are not commonly trained on the Edge since the cost of transferring training data to the edge node is much higher than just transferring a trained model. This does not apply when the data is collected on the node itself, and automatically labelled so it can be used in training. We discuss such a scenario in Section~\ref{sec:viewpoint}. 
Even if we can get enough prior training data onto the Edge node and any additional data being captured on the node is only relevant to training the current node, there is still the issue of limited computational capabilities of an Edge node. We detail this problem in Section~\ref{sec:memory_issues} and solutions in Section~\ref{sec:checkpointing}.

\section{Array of things and the Waggle platform}
\label{sec:waggle}
{\em Array of Things} is an Internet-of-Things project that uses an array of hundreds of sensors that work to collect data as a single unit, much like a telescope array, but with sensors collecting data about a city instead. Under this project, hundreds of smart-sensor nodes have been placed all over the city of Chicago. These nodes integrate air-quality sensors with cameras and on-board computational capability to create a distributed and integrated, city-wide network of smart-sensors that can be programmed and controlled as a single instrument to capture data for scientific research~\cite{catlett2017array}.

The individual Edge nodes of this network are based on the Waggle platform~\cite{beckman2016waggle}\footnote{http://wa8.gl/}, which is designed as an embedded system with sensing and edge computing capabilities. It packages sensors for environmental measurements like pressure, temperature, humidity, light (IR/UV), sound level, gas levels, along with a camera, and three single-board computers (SBCs) into one node. Only one of these SBCs is meant to run edge computing payload, while the others are for node management and reliability. The current payload SBC is an ODROID XU4 based on the Samsung Exynos5422 CPU with 
four  A15  cores,  four  A7  cores,  a  Mali-T628  MP6 GPU  that  supports  OpenCL, 2GB  LPDDR3 RAM, and attached flash storage. Due to the limited computational capacity of the Odroid platform, other options are being considered. These nodes are already running OpenCV, Caffe and Tensorflow.

With a camera and on-board computational capabilities, an obvious use is to run visual machine learning models, for example, to count the number of people in an area to understand the usage of streets and public spaces, or the number and types of cars passing by. This platform is also being used for flood and ice detection. 
All these applications, however, currently only do inference on these Waggle nodes and no training. Some of the reasons for this have been explained in Section~\ref{sec:introduction}. However, most of these visual models suffer from the viewpoint problem where the images on which these models are operating are from an angle which is specific to the installation of the particular camera/node (see Section~\ref{sec:viewpoint}).

\section{The viewpoint problem and in-situ student-teacher training}
\label{sec:viewpoint}

The viewpoint problem is a common problem in computer vision, faced when training an image classification or segmentation model on a data set. One example is a face recognition model. If all the facial images used to train the model are taken at eye level and with the subject directly facing the camera, the model will be trained to recognize faces in images taken at similar angles only and may not be effective on images taken from different angles. This model suffers from the viewpoint problem if used directly. In this approach, we use it as a ``teacher" model instead. 

In the context of the Waggle platform, although this teacher model may not be able to detect faces at certain skewed angles, it may still work at other angles that are closer to the original training angle. For example, let us assume that a subject walks from the left to the right edges of the frame, and the teacher model correctly identifies it in the last frame. Having received this identification, an object-tracking model \cite{brunetti2018computer} can be used to identify and segment all the previous frames which contain the same subject.
These frames are then set aside, along with the identified label, as part of a new training set. Every such instance of the teacher model identifying a subject contributes tens of images to this new dataset, which can then be used to train a new ``student'' model. This student-teacher paradigm has previously been used to compress large networks into more parameter-efficient networks. \cite{crowley2018moonshine}

Since these images will be used to train a convolutional neural network, they do not need to be stored at a very high resolution. At the standard resolution of $224 \times 224$, the size can be expected to be less than 10kb per image. Storing even about 100,000 of these images would require about 10GB of local storage, which is easily provided on an SD card - present on Waggle nodes. Since the training of the student model is not time critical, it can then be scheduled to run only when the node's CPU does not have a higher priority task. 
Doing this, we can specialise the model running at each node to its own viewpoint, automatically improving its accuracy. 
\label{sec:memory_issues}

Although in-situ training might be useful to address the viewpoint problem, there are computational issues - specifically, vision models typically require a large amount of memory. Tables~\ref{mem:fixed_image},~\ref{mem:fixed_batch}~and~\ref{mem:fixed_batch8} detail the amount of memory required for the whole model and activations for a few variations of ResNet, since that is a popular model for vision problems.

It can be seen in Table~\ref{mem:fixed_image} that all models fit in 2GB memory, without taking into account the memory needed to perform computations. However, increasing the batch size to 3 makes it impossible to keep ResNet152 in memory and further increase makes even the smallest models require more than 2GB. Since training models using extremely small batch sizes is inefficient because of a large number of minibatches per epoch \cite{keskar2016large}, finding a way to increase the batch size while keeping the model in memory can improve training performance. 

At the same time problems with memory could also emerge for images with higher resolution than the standard one ($224 \times 224$) as it follows from Table~\ref{mem:fixed_batch}, even for the smallest batch\_size. The situation becomes worse for batch\_size = 8, when one cannot use a neural network with more than 50 layers even for the smallest possible image size. 
\begin{table}
\begin{tabular}[h]{|l|c|c|c|c|c|}
\hline
& \multicolumn{5}{c|}{ResNet$_x$}\\
\hline
batch\_size & $x = 18$ & $x = 34$ & $x = 50$ & $x = 101$ & $x = 152$ \\
\hline
 1&230.05 &413.00 &620.27 & 1027.21& 1410.62\\
 3&340.05&580.42&1091.11&1732.33& \cellcolor{Gray}2405.14\\
 5& 450.06& 747.85 & 1561.94&\cellcolor{Gray} 2437.45& \cellcolor{Gray}3399.67\\
 10 & 725.07& 1166.42 &\cellcolor{Gray} 2739.04 &\cellcolor{Gray} 4200.25& \cellcolor{Gray} 5885.98\\
 30 & 1825.13&\cellcolor{Gray} 2840.70& \cellcolor{Gray} 7447.42&\cellcolor{Gray} 11251.43&\cellcolor{Gray} 15831.23 \\
 50 &\cellcolor{Gray} 2925.18 &\cellcolor{Gray} 4514.97 &\cellcolor{Gray} 12155.79&\cellcolor{Gray} 18302.62&\cellcolor{Gray} 25776.48\\

\hline
\end{tabular}
\caption{Memory requirement for each model to keep all weights and activations for the standard size of image ($224 \times 224$), the amount is  given in MB. The shaded values correspond to the cases when the model cannot fit in memory.}
\label{mem:fixed_image}
\end{table}

\begin{table}
\begin{tabular}[h]{|p{13mm}|c|c|c|c|c|}
\hline

& \multicolumn{5}{c|}{ResNet$_x$}\\
\hline
image width/height& $x = 18$ & $x = 34$ & $x = 50$ & $x = 101$ & $x = 152$ \\
\hline
 224&230.05 &413.00 &620.27 & 1027.21& 1410.62\\
 350&309.83&534.96&964.66&1543.72&\cellcolor{Gray}2139.75\\
 500& 449.21& 749.73 & 1570.93&\cellcolor{Gray} 2472.72& \cellcolor{Gray}3458.50\\
 650 & 639.07& 1039.08 &\cellcolor{Gray} 2387.54 &\cellcolor{Gray} 3682.00& \cellcolor{Gray} 5161.76\\
 1100 & 1496.10&\cellcolor{Gray} 2346.95&\cellcolor{Gray} 6073.06&\cellcolor{Gray} 9208.30&\cellcolor{Gray} 12961.96 \\
 1500 &\cellcolor{Gray} 2628.70 &\cellcolor{Gray} 4075.07 &\cellcolor{Gray} 10944.42&\cellcolor{Gray} 16515.11&\cellcolor{Gray} 23277.27\\

\hline
\end{tabular}

\caption{Memory requirement for each model to keep all weights and activations for the batch\_size = 1, the amount is given in MB. The shaded values correspond to the cases where the model cannot fit in memory.}
\label{mem:fixed_batch}
\end{table}

\begin{table}
\begin{tabular}[h]{|p{13mm}|c|c|c|c|c|}
\hline
& \multicolumn{5}{c|}{ResNet$_x$}\\
\hline
image width/height& $x = 18$ & $x = 34$ & $x = 50$ & $x = 101$ & $x = 152$ \\
\hline
 224&0.60 &0.98&\cellcolor{Gray}2.22& \cellcolor{Gray}3.41& \cellcolor{Gray}4.78\\
 350&1.22&1.93&\cellcolor{Gray}4.90&\cellcolor{Gray}7.45&\cellcolor{Gray}10.47\\
 500& \cellcolor{Gray}2.31& \cellcolor{Gray}3.60& \cellcolor{Gray}9.63&\cellcolor{Gray} 14.69&\cellcolor{Gray} 20.76\\
 650 &\cellcolor{Gray} 3.79&\cellcolor{Gray} 5.86&\cellcolor{Gray} 15.99 & \cellcolor{Gray}24.13& \cellcolor{Gray}34.06\\

\hline
\end{tabular}

\caption{Memory requirement for each model to keep all weights and activations for the batch\_size = 8, the amount is given in GB. The shaded values correspond to the cases where the model cannot fit in memory.}
\label{mem:fixed_batch8}
\end{table}


\section{Related work on reducing memory consumption for backpropagation}
\label{sec:checkpointing}

Training a neural network through backpropagation has a characteristic data flow pattern where the data corresponding to the neuron activations is generated while propagating forward through the network. This is followed by a backpropagation pass that calculates derivatives, using the activation values computed earlier. This means that a simple implementation of backpropagation would require all the activations to be stored in memory during the forward pass, in order to use them again during backpropagation. 

Sometimes, the memory required to do this is not available. Memory issues are not uncommon, even when training on the largest available commodity GPUs today - the batch size is often adjusted so that a single batch can fit in memory - however the batch size also affects the convergence properties of the training. This problem is especially intensified when training on the Edge when the available batch sizes might be as low as 1-2 (if at all). \cite{hanlon_2018}
In recent times, multiple studies have addressed these memory concerns. One technique is model parallelism, where a big model is split over multiple nodes in a cluster \cite{wang2018supporting, huang2018gpipe}. However this is only applicable when the training is done on a cluster with a high-speed interconnect as otherwise communication overheads quickly dominate. 

Another technique that has garnered attention recently is checkpointing, also sometimes called reforwarding. In this approach, only a subset of activations is stored during the forward pass, and the rest discarded. The discarded data can be recovered by rerunning the forward propagation from the last available ``checkpoint''. \cite{feng2018cutting}
Although most major neural network training packages today have some implementation of checkpointing \cite{chen2016training, gruslys2016memory, chen2015mxnet} these implementations are very basic and do not take advantage of the research that was done on this topic in the fields of high performance computing and automatic differentiation. This means that some of these implementations might be doing more computations or using more memory than strictly necessary. We discuss this in Section~\ref{sec:checkpointingbenefits}. 
While the suboptimality of these implementations might not be immediately obvious when training on a cluster, more efficient implementations are required when dealing with hardware that is highly limited in its computational powers - not only does it not have enough memory to store the entire model, its CPU is small enough that the effect of suboptimal recomputation will be more obvious.

\section{Existing checkpointing implementations in machine learning frameworks}


PyTorch is a fast-evolving Python package widely applied in deep learning. It is developed by Facebook's artificial-intelligence research group. It shares some features with another popular package called TensorFlow. Both use Tensors as a basic class and all operations are performed on them. The structure of Tensors is similar to the one used in NumPy library with the same basic functions and operations, while also allowing to work with them on GPU. A key difference is Pytorch's ability to dynamically define the computational graph of the model. That renders models flexible, allowing them to be changed during training. Additionally, PyTorch is considered more transparent and more Python oriented.

PyTorch is actively being developed and is designed to be memory efficient to allow executing larger models. To achieve this memory efficiency several techniques are applied, including data-flow analysis, data parallelism and checkpointing, as discussed in Section~\ref{sec:checkpointing}. The current implementation called \textit{checkpoint\_sequential}  divides the whole network in parts that are all equal except the last one. The number of such parts is determined by the parameter \textit{segments}. Hence, during the forward propagation phase only the inputs of first segments are saved for the backward propagation phase and the last segment is treated as usual, i.e. all activations are stored. So during the backward phase the last segment could be processed immediately while for others it is necessary to recompute activations starting from checkpoints in order to proceed.

For better understanding of how much memory consumption could be reduced it is enough to consider the following example. Assuming that there is a neural network which could be divided in $l$ homogeneous blocks in terms of operation costs and activation sizes and $s$ corresponds to the number of segments described above, then the total memory taken by all activations could be defined by
$$
\mbox{Memory} = s - 1 + \bigl(l - \left\lfloor l/s \right\rfloor (s -1) \bigr).
$$
It can be seen that there is a lower bound $L = 2\sqrt{l}$ and for any $s \geq 2$ it is not possible to reduce $Memory$ below this value. On the other hand, for Edge devices it can be crucial if large models are used (see Section~\ref{sec:memory_issues}). Therefore, in case of bigger models another approach should be applied like binomial checkpointing discussed in more details  in Section~\ref{sec:checkpointingbenefits}.



\section{Proposed Improvements}
\label{sec:checkpointingbenefits}


In this section, we evaluate the practical advantages of optimal checkpointing. 

In order to establish this, we will base our analysis on the memory requirements of different ResNet networks, different batch size and different image sizes, as given in Tables~\ref{mem:fixed_image} and~\ref{mem:fixed_batch}. To simplify the analysis here, we will denote by LinearResNet$_x$ a linear homogeneous network built by analogy to ResNet$_x$. The memory needed to store all network weights is the same in LinearResNet$_x$ and in ResNet$_x$, and the size of the forward activation for a given image size in LinearResNet$_x$ is defined as the overall activation weights for ResNet$_x$ divided by the depth of ResNet$_x$. Thus, we obtain a linear homogeneous version LinearResNet$_x$ of ResNet$_x$, that has approximately the same memory requirements as ResNet$_x$, both in terms of weights and activations. 

Considering a LinearResNet of depth $l$, let us denote by $M_{C}$ the size of the memory of the Edge device, by $M_{W}$ the memory needed to store all the weights, and by $M_{A}$ the memory to store the result of any forward step (remember that all the layers in the LinearResNet have the same size) with a unit batch size, and let us denote by $k$ the batch size. Then, $n_{\max}=\frac{M_{C}-M_{W}}{k \times M_{A}}$ represents the depth of the largest ResNet that can be trained in the given amount of memory for the associated batch size $k$. In order to increase $n_{\max}$, the solution that is used in practice often consists in using a smaller batch size $k$, which may affect the time to complete an epoch. On the other hand, for $n_{\max}$ greater than 3, it is possible to rely on checkpointing in order to perform training. 
In fact, given $n_{\max}$, it is possible to determine in polynomial time the optimal dynamic sequence of checkpoints, using the dynamic programming algorithm Revolve~\cite{revolve, pyrevolve,code_julien}. The produced computation schedule is recursive, in the sense that the same memory slot is used to store activations from different layers at different times. We refer the reader to~\cite{pyrevolve} for the details of the algorithm, and our goal in this section is rather to show that optimal checkpointing is very efficient in drastically limiting required memory, while only reasonably increasing the processing time.

\begin{figure}[t!]
    \hspace{-0.5cm}\begin{subfigure}[b]{0.25\textwidth}
    \includegraphics[width=5.5cm]{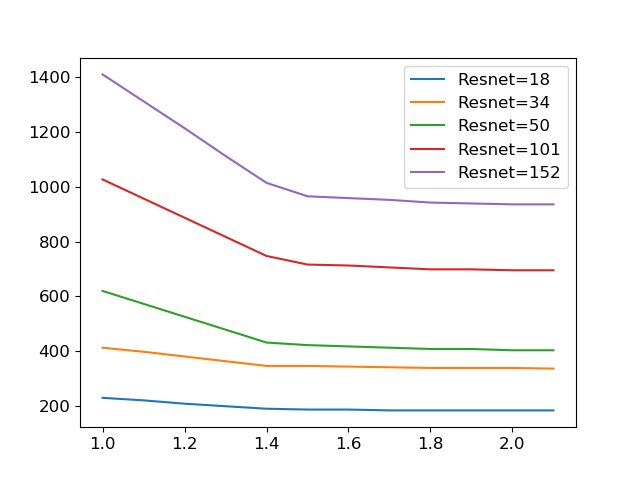}
    \caption{Batch Size 1, Image Size 224}
    \label{f224}
    \end{subfigure}~~~~
    \begin{subfigure}[b]{0.25\textwidth}
    \includegraphics[width=5.5cm]{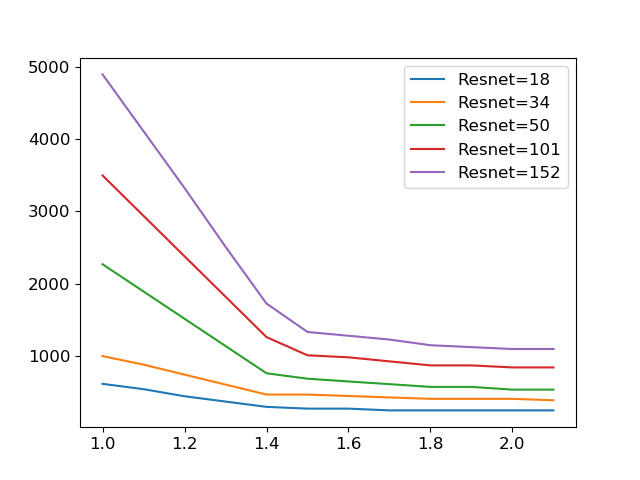}
    \caption{Batch Size 8, Image Size 224}
    \label{8f224}
    \end{subfigure}
    
    \hspace{-0.5cm}\begin{subfigure}[b]{0.25\textwidth}
    \includegraphics[width=5.5cm]{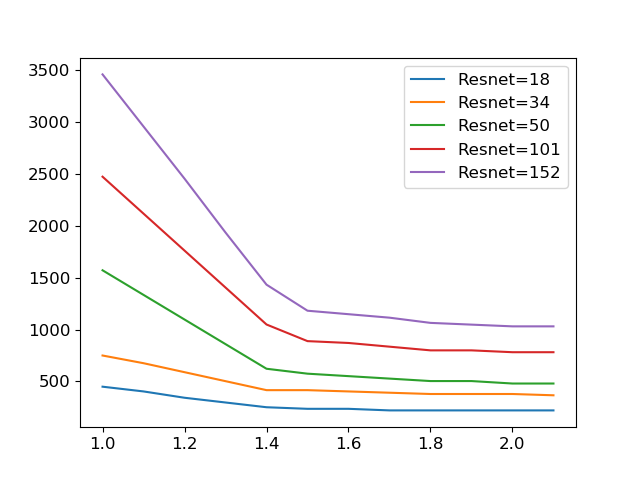}
    \caption{Batch Size 1, Image Size 500}
    \label{f500}
    \end{subfigure}~~~~
    \begin{subfigure}[b]{0.25\textwidth}
    \includegraphics[width=5.5cm]{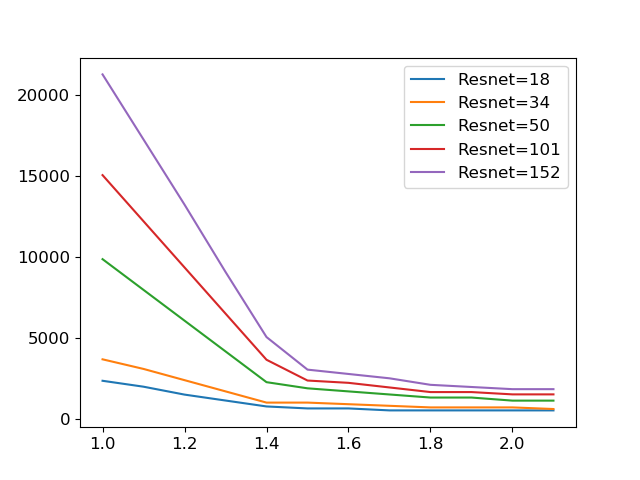}
    \caption{Batch Size 8, Image Size 500}
    \label{8f500}
    \end{subfigure}%
    \caption{Peak memory requirement vs recomputate factor for different image sizes. The recompute factor is the ratio between the extended time to solution due to recomputations induced by the memory-saving checkpointing and the original time to solution. We can see that for small recompute factors, the memory requirement is often prohibitively high, especially for an Edge node}
    \label{plots}
\end{figure}

Let us denote by $\rho$ the increase in the time for a single backpropagation, i.e. the recompute factor, that is acceptable in our specific context. Thus, the maximal number of forward and backward computations is $2 \rho l$. Combining PyRevolve, that computes the optimal schedule that minimizes the time to solution given a number of checkpoint slots to an elementary binary search, we can easily compute the minimal number of checkpoint slots so that the time to solution is smaller than $2 \rho l$. Figure~\ref{plots} depict the evolution of the memory footprint with $\rho$ for different LinearResNet networks. Each plot corresponds to a specific image size, either small ($224 \times 224$) for a respective batch size of 1 (Figures~\ref{f224}) and 8 (Figures~\ref{8f224})and medium ($500 \times 500$) for a respective batch size of 1 (Figures~\ref{f500}) and 8 (Figures~\ref{8f500}).

$\rho=1$ corresponds to the case with no checkpointing. In this specific case, we can observe that all models and activations can fit into the 2GB limit only if the image size is 224. In all other cases (larger batch sizes or larger image sizes), the memory is too limited to store and run the models. 
On the other hand, considering a value of $\rho$ between $1.5$ and $2$ dramatically changes the situation. The lower memory consumption can be used to consider larger batch sizes. For instance, when the batch size is 8 (Figure~\ref{8f500}), all models fit into the 2GB memory with $\rho>1.6$, whereas in the same context, even ResNet$_{18}$ does not fit into the 2GB limit.

Moreover, the effective increase in the total time to solution is likely to be smaller than what is shown in above results because a larger batch size will enable fewer batches per epoch. Also, on the typical multi-threaded vector architectures (such as GPUs), having a larger batch-size enables to increase the computational efficiency, and therefore, the time to process 8 times a batch size of 1 is expected to be much larger than the time to process a batch size of $8$.

\section{Conclusion}
We have considered the opportunity of performing training on Edge devices, especially in the context of the viewpoint problem. A student-teacher model pair is a possible approach whereby the viewpoint problem can be addressed by in-situ training of a model specialized to each camera's viewpoint. This approach does not require any additional data to be transferred to the node beyond the original teacher model. We have also shown that the peak memory footprint, which is a crucial factor for training on Edge devices, can be reduced by checkpointing strategies. We show that the current implementations of checkpointing in popular neural network packages can be improved by taking advantage of full binomial checkpointing and that the impact of this improvement would be most useful for training on the Edge. 

{\bf Acknowledgments}:
This paper benefited greatly from discussions with Paul Kelly and Prasanna Balaprakash. This work was partly funded by the Intel Parallel Computing Centre at Imperial College London and EPSRC EP/R029423/1. This work was partly funded by HPC-BigData INRIA Project LAB (IPL).This work was funded in part by a grant from U.S. Department of Energy,
Office of Science,  under contract DE-AC02-06CH1135.

\bibliographystyle{abbrv}
\bibliography{references}

\scriptsize
\framebox{\parbox{4.8in}{The submitted manuscript has been created by UChicago Argonne, LLC, Operator of Argonne National Laboratory (`Argonne'). Argonne, a U.S. Department of Energy Office of Science laboratory, is operated under Contract No. DE-AC02-06CH11357. The U.S. Government retains for itself, and others acting on its behalf, a paid-up nonexclusive, irrevocable worldwide license in said article to reproduce, prepare derivative works, distribute copies to the public, and perform publicly and display publicly, by or on behalf of the Government.  The Department of Energy will provide public access to these results of federally sponsored research in accordance with the DOE Public Access Plan. \url{http://energy.gov/downloads/doe-public-access-plan}.}}
\normalsize
\end{document}